\documentclass[a4paper,11pt]{article}
\pdfoutput=1
\usepackage{jheppub_mod} 
\usepackage[T1]{fontenc} 

\usepackage{hyperref}
\usepackage{graphicx}
\usepackage{amsmath,amssymb,slashed}
\usepackage{booktabs,tabulary}
\usepackage{dsfont}
\usepackage{color}
\usepackage{multirow}
\usepackage{subcaption}

\newcommand{\Order}{\mathcal{O}}

\newcommand{\GeV}{\,\text{GeV}}
\newcommand{\TeV}{\,\text{TeV}}

\newcommand{\Br}{\text{Br}}
\newcommand{\Lagr}{\mathcal{L}}
\newcommand{\beq}{\begin{equation}}
\newcommand{\eeq}{\end{equation}}
\renewcommand{\Re}{\text{Re}\,}

\allowdisplaybreaks[1]

\preprint{CERN-TH-2021-050, PSI-PR-21-04, ZU-TH 14/21}
\title{Consequences of chirally enhanced explanations of $\boldsymbol{(g-2)_\mu}$ for $\boldsymbol{h\to \mu\mu}$ and $\boldsymbol{Z\to \mu\mu}$}

\author[a,b,c]{Andreas Crivellin}
\author[d]{and Martin Hoferichter}

\affiliation[a]{
CERN Theory Division, CH--1211 Geneva 23, Switzerland}

\affiliation[b]{
Paul Scherrer Institut, CH--5232 Villigen PSI, Switzerland}

\affiliation[c]{
Physik-Institut, Universit\"at Z\"urich, Winterthurerstrasse 190, CH--8057 Z\"urich, Switzerland}

\affiliation[d]{Albert Einstein Center for Fundamental Physics, Institute for Theoretical Physics, University of Bern, Sidlerstrasse 5, CH--3012 Bern, Switzerland}

\emailAdd{andreas.crivellin@cern.ch}
\emailAdd{hoferichter@itp.unibe.ch}

\abstract{
With the long-standing tension between experiment and Standard-Model (SM) prediction in the anomalous magnetic moment of the muon $a_\mu$ recently reaffirmed by the Fermilab experiment, the crucial question becomes which other observables could be sensitive to the underlying physics beyond the SM to which $a_\mu$ may be pointing.
While from the effective field theory (EFT) point of view no direct correlations exist, this changes in specific new physics models. In particular, in the case of explanations involving heavy new particles above the electroweak (EW) scale with chiral enhancement, which are preferred to evade exclusion limits from direct searches, correlations with other observables sensitive to EW symmetry breaking are expected. Such scenarios can be classified according to the $SU(2)_L$  representations and the hypercharges of the new particles. We match the resulting class of models with heavy new scalars and fermions onto SMEFT and study the resulting correlations with $h\to\mu\mu$ and $Z\to\mu\mu$ decays, where, via $SU(2)_L$ symmetry, the latter process is related to $Z\to\nu\nu$ and modified $W$--$\mu$--$\nu$ couplings. 
}

\begin{document} 
\maketitle

\section{Introduction}

The recent release of Run 1 data of the Fermilab Muon $g-2$ experiment~\cite{Abi:2021gix,Albahri:2021ixb,Albahri:2021kmg,Albahri:2021mtf} confirms the previous measurement at Brookhaven National Laboratory~\cite{Bennett:2006fi}, leading to a new combined world average of 
\begin{equation}
 a_\mu^\text{exp}=116\,592\,061(41)\times 10^{-11},
\end{equation}
which differs from the SM theory prediction~\cite{Aoyama:2020ynm,Aoyama:2012wk,Aoyama:2019ryr,Czarnecki:2002nt,Gnendiger:2013pva,Davier:2017zfy,Keshavarzi:2018mgv,Colangelo:2018mtw,Hoferichter:2019gzf,Davier:2019can,Keshavarzi:2019abf,Hoid:2020xjs,Kurz:2014wya,Melnikov:2003xd,Colangelo:2014dfa,Colangelo:2014pva,Colangelo:2015ama,Masjuan:2017tvw,Colangelo:2017qdm,Colangelo:2017fiz,Hoferichter:2018dmo,Hoferichter:2018kwz,Gerardin:2019vio,Bijnens:2019ghy,Colangelo:2019lpu,Colangelo:2019uex,Blum:2019ugy,Colangelo:2014qya}\footnote{Some recent developments include Refs.~\cite{Borsanyi:2020mff,Lehner:2020crt,Crivellin:2020zul,Achasov:2020iys,Keshavarzi:2020bfy,Malaescu:2020zuc,Colangelo:2020lcg} for hadronic vacuum polarization and Refs.~\cite{Hoferichter:2020lap,Ludtke:2020moa,Bijnens:2020xnl,Bijnens:2021jqo,Zanke:2021wiq,Chao:2021tvp,Danilkin:2021icn,Colangelo:2021nkr} for hadronic light-by-light scattering.}  
\begin{equation}
\label{amuSM}
a_\mu^\text{SM}=116\,591\,810(43)\times 10^{-11}
\end{equation}
by $4.2\sigma$. If this tension indeed signals physics beyond the SM (BSM), the most pressing challenge becomes unraveling its nature. Given that the difference $\Delta a_\mu=a_\mu^\text{exp}-a_\mu^\text{SM}=251(59)\times 10^{-11}$ is even larger than the EW contribution, $a_\mu^\text{EW}=153.6(1.0)\times 10^{-11}$~\cite{Czarnecki:2002nt,Gnendiger:2013pva}, any BSM explanation with new particles needs to invoke some enhancement mechanism. A promising class of solutions achieves this by avoiding a SM-like scaling $a_\mu^\text{BSM}\propto m_\mu^2$, 
with the chirality flip originating from a large coupling to the SM Higgs
instead of the small muon Yukawa coupling in the SM. 
This chiral enhancement allows for viable solutions for particle masses up to tens of TeV~\cite{Czarnecki:2001pv,Stockinger:1900zz,Giudice:2012ms,Altmannshofer:2016oaq,Kowalska:2017iqv,Crivellin:2018qmi,Capdevilla:2021rwo}. 

Explicit models that realize this mechanism include   
the minimal supersymmetric SM (MSSM), where the enhancement factor is provided by $\tan \beta$~\cite{Moroi:1995yh,Feng:2001tr,Stockinger:2006zn}, taking values around $50$~\cite{Ananthanarayan:1991xp,Carena:1994bv} in the case of top--bottom Yukawa coupling unification. For universal supersymmetry breaking mechanisms the LHC bounds on the supersymmetric partners are already so stringent that this enhancement~\cite{Lopez:1993vi,Chattopadhyay:1995ae,Dedes:2001fv} is insufficient to explain $a_\mu$, but less minimal scenarios can still work~\cite{Bagnaschi:2017tru}.
Next, leptoquark (LQ) models can even display an enhancement factor of $m_t/m_\mu \approx 1700$~\cite{Djouadi:1989md, Chakraverty:2001yg,Cheung:2001ip,Bauer:2015knc,ColuccioLeskow:2016dox,Crivellin:2020tsz,Crivellin:2020mjs,Fajfer:2021cxa,Greljo:2021xmg}, allowing for a TeV-scale explanation with perturbative couplings that evades direct LHC searches and might even explain the other anomalies pointing towards lepton flavor universality violation. Other models in which chiral enhancement may be present 
include composite or extra-dimensional models~\cite{Das:2001it,Xiong:2001rt,Park:2001uc}, models with two Higgs doublets~\cite{Iltan:2001nk,Broggio:2014mna,Crivellin:2015hha,Crivellin:2019dun}, or $Z^\prime$ models with $\tau\mu$ couplings~\cite{Altmannshofer:2016brv,Buras:2021btx}.

In this work, we study the question which correlations exist with other processes, specifically $h\to\mu\mu$ and $Z\to\mu\mu$ decays, if indeed such a BSM scenario with chiral enhancement is realized. For this purpose, we consider a class of models with new scalars and fermions that display the minimal features to implement the chiral enhancement, allowing for a wide range of $SU(2)_L$ representations and hypercharges. As a first step we match these models onto the relevant set of dimension-$6$ effective operators in SMEFT~\cite{Buchmuller:1985jz,Grzadkowski:2010es}, based on which correlations have been pointed out in Refs.~\cite{Buttazzo:2020eyl,Yin:2020afe,Aebischer:2021uvt}. However, this assumes that only a single or a few Wilson coefficient are non-zero at the matching scale, while if the full set of possible initial conditions is taken into account, the size of the SMEFT parameter space is not reduced.  The identification of correlations beyond SMEFT relations originating from $SU(2)_L$ invariance is only possible with additional assumptions, such as implemented in simplified models. 

In order to achieve chiral enhancement in $a_\mu$ with new particles in the loop, at least three fields (two scalars and one fermion or two fermions and one scalar) are needed, some of which, as, e.g., in LQ models, can be taken from the SM. As a first step, 
we will classify the possible representations of three new fields under $SU(2)_L\times U(1)_Y$ and perform the matching onto the relevant SMEFT operators. This allow us to identify the correlations among different observables, which we then study numerically for some selected cases. 

\section{Simplified models and EFT analysis}
\label{sec:models}

\begin{table}[t!]
	\centering
	\begin{tabular}{cc|ccc|ccc}
		\toprule
		& $R$ & ${\Psi ,\Phi }$&${{\Phi _L},{\Psi _L}}$&${{\Phi _E},{\Psi _E}}$ & $\phi$ & $\ell$ & $e$\\\midrule
		\multirow{4}{*}{$SU(2)_L$} & $121$ & $1$ & $2$ & $1$ & \multirow{4}{*}{$2$} & \multirow{4}{*}{$2$} & \multirow{4}{*}{$1$}\\
		& $212$ & $2$ & $1$ & $2$ & & &\\
		& $323$ & $3$ & $2$ & $3$ & & &\\
		& $232$ & $2$ & $3$ & $2$ & & &\\
		$Y$ & & $X$ & $-\frac{1}{2}-X$ & $-1-X$ & $\frac{1}{2}$ & $-\frac{1}{2}$ &$-1$\\
		\bottomrule
	\end{tabular}
	\caption{Charge assignments and representations under $SU(2)_L\times U(1)_Y$ for the SM fields and the different new particles.}
	\label{quantum_numbers}
\end{table}

\begin{table}[t]
	\centering
	\begin{tabular}{crrrr}
		\toprule
		$R$& $121$ & $212$ & $323$ & $232$\\\midrule
		$\xi_{e\phi}$ & $1$ & $-1$ & $5$ & $5$\\
		$\xi_{eB}$ & $1$ & $-1$ & $3$ & $3$\\
		$\xi_{eW}^L$ & $\frac{1}{2}$ & $0$ & $-\frac{1}{2}$ & $2$\\
		$\xi_{eW}^E$ & $0$ & $\frac{1}{2}$ & $-2$ & $\frac{1}{2}$\\
		$\tilde\xi_{eW}$ & $0$ & $-\frac{1}{2}$ & $2$ & $-\frac{1}{2}$\\
		$\xi_{\phi e}$ & $1$ & $1$ & $3$ & $3$\\
		$\xi_{\phi e}'$ & $1$ & $2$ & $3$ & $2$\\
		$\xi_{\phi \ell}^{(1)}$ & $-\frac{1}{2}$ & $-1$ & $-\frac{9}{2}$ & $-3$\\
		$\xi_{\phi \ell}'^{(1)}$ & $1$ & $1$ & $3$ & $3$\\
		$\xi_{\phi \ell}^{(3)}$ & $-\frac{1}{2}$ & $0$ & $-\frac{1}{2}$ & $-2$\\
		$\xi_{\phi\ell}'^{(3)}$ & $\frac{1}{4}$ & $0$ & $-\frac{1}{4}$ & $1$\\
		$\xi_{\phi \ell}''^{(3)}$ & $0$ & $\frac{1}{4}$ & $1$ & $-\frac{1}{4}$\\
		\bottomrule
	\end{tabular}
	\caption{Representation-dependent $SU(2)_L$ factors $\xi$ entering the matching calculation.}
	\label{xi_R} 
\end{table}

There are two classes of models that display chiral enhancement for $a_\mu$: (I)  two scalars $\Phi_{L,E}$ and one fermion $\Psi$ and  
(II) two fermions $\Psi_{L,E}$ and one scalar $\Phi$.\footnote{Similar simplified setups have also been considered in the context of $b\to s\ell^+\ell^-$~\cite{Gripaios:2015gra,Arnan:2016cpy,Arnan:2019uhr} and radiative muon mass models~\cite{Kannike:2011ng,Dermisek:2013gta,Freitas:2014pua,Baker:2021yli}. Note that also new fermion and vector particles can display chiral enhancement. However, massive vectors are not renormalizable without a Higgs mechanism, so that the effects in $h\to\mu\mu$ and $Z\to\mu\mu$ cannot be calculated in a simplified setup, see also Ref.~\cite{Brod:2019bro}. Our analysis covers the LQ models S1 and S2 in case II with $R=121$ and $212$, respectively, and upon identifying $\Psi_L=t_L$, $\Psi_E=t_R$, and  $\kappa=Y_t$.} We define the Lagrangians in these two cases as
\begin{align}
{\Lagr_\text{I}} &= \lambda _L^{\text{I}}\,\bar \ell\Psi {\Phi _L} + \lambda _E^{\text{I}}\,\bar e\Psi {\Phi _E} + A\,\Phi _L^\dag \Phi _E^{}\phi,\notag\\
{\Lagr_\text{II}} &= \lambda _L^\text{II}\,\bar \ell{\Psi _L}\Phi  + \lambda _E^\text{II}\,\bar e{\Psi _E}\Phi  + \kappa\, {{\bar \Psi }_L}{\Psi _E}\phi,
\end{align}
where $\ell$, $e$, and $\phi$ are the lepton doublet, singlet, and Higgs field of the SM (throughout, we follow the notation of Ref.~\cite{Grzadkowski:2010es}). The conventions for the SM particles and the $SU(2)_L$ quantum numbers and hypercharges $Y$ are given in Table~\ref{quantum_numbers}. We consider the four combinations of $SU(2)_L$ representations ($R$) up-to-and-including triplets (see also Ref.~\cite{Calibbi:2018rzv}), while the hypercharge assignment can be parameterized in terms of a general variable $X$. The details of the $SU(2)_L$ contractions in the various cases are spelled out in Eq.~\eqref{SU2contractions}. We further assume a $Z_2$ symmetry to avoid mixing with SM fields, which could generate tree-level effects in $h\to\mu\mu$ and $Z\to\mu\mu$ that are at least strongly disfavored~\cite{ALEPH:2005ab,Crivellin:2018qmi,Zyla:2020zbs,Crivellin:2020ebi,Aad:2020xfq,Sirunyan:2020two,Crivellin:2020tsz}.\footnote{Note, however, that these bounds can be relaxed within 2HDMs~\cite{Dermisek:2021ajd,Dermisek:2020cod}.} Moreover, $a_\mu$ is only generated at loop level, further motivating the study of loop effects in $h\to\mu\mu$ and $Z\to\mu\mu$ as well.

\begin{figure}[t]
\centering
	\includegraphics[width=0.75\linewidth]{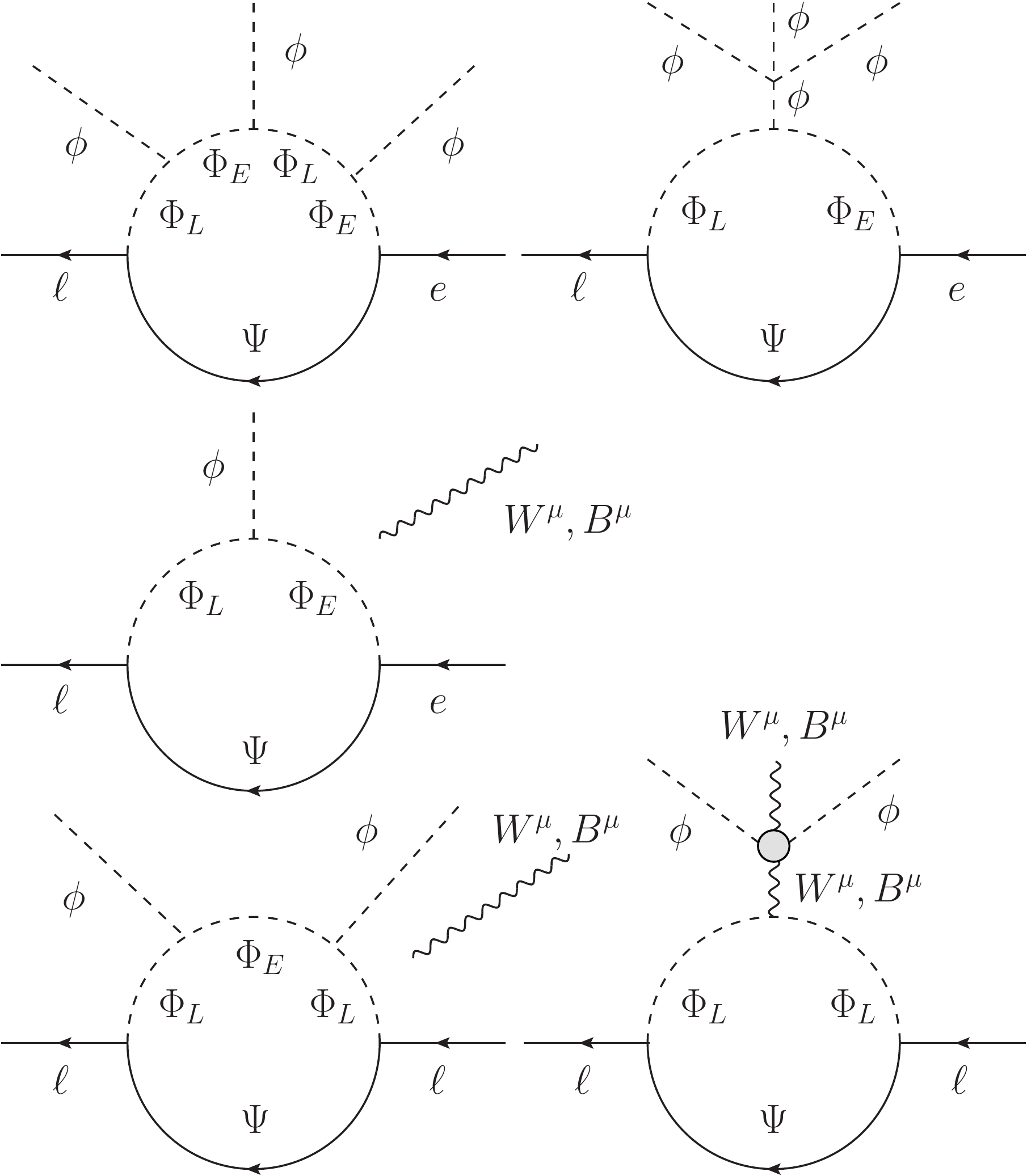}
	\caption{Diagrams contributing to the matching of the simplified models onto the SMEFT operators~\eqref{EFT_operators}. The first line is relevant for $Q_{e\phi}$, the second for $Q_{eB}$, $Q_{eW}$, and the third for $Q_{\phi \ell }^{\left( 1 \right)}$, $Q_{\phi \ell }^{\left( 3 \right)}$ (the diagrams for $Q_{\phi e}$ follow from $\ell \to e$ and $L\leftrightarrow E$). In the third and fourth diagram the gauge boson is understood to couple at all possible places, while the gray blob in the fifth diagram denotes the full tree-level scalar QED amplitude. Note that after $SU(2)_L$ breaking the second and fifth diagram only contribute to on-shell decays but not to effective $h$, $W$, and $Z$ couplings at zero momentum transfer.} 
	\label{fig:diagrams}
\end{figure}

The effective operators in the SMEFT Lagrangian that directly generate $(g-2)_\mu$, $h\to\mu\mu$, $Z\to\mu\mu,\nu\nu$,  and $W\to\mu\nu$ (after EW symmetry breaking) are~\cite{Grzadkowski:2010es}
\begin{align}
Q_{e\phi } &= {\phi ^\dag }\phi\, \bar \ell e\phi,
& Q_{\phi e} &= {\phi ^\dag }i\overset\leftrightarrow{D}^\mu\phi \,\bar e{\gamma _\mu }e,\notag\\
{Q_{eB}} &= \bar \ell {\sigma _{\mu \nu }}e\phi\, B^{\mu \nu },&
Q_{\phi \ell }^{\left( 1 \right)} &= {\phi ^\dag }i\overset\leftrightarrow{D}^\mu
\phi\, \bar \ell {\gamma _\mu }\ell,\notag\\
{Q_{eW}} &= \bar \ell {\sigma _{\mu \nu }}e{\tau ^I}\phi\, W_I^{\mu \nu },&
Q_{\phi \ell }^{\left( 3 \right)} &= {\phi ^\dag }i\overset\leftrightarrow{D}_I^\mu \phi\, \bar \ell \tau^I{\gamma _\mu }\ell,
\label{EFT_operators}
\end{align}
and the relevant diagrams that contribute to the matching are shown in Fig.~\ref{fig:diagrams}, which we calculate in the limit of a vanishing muon Yukawa coupling as it cannot lead to chiral enhancement.

In the equal-mass case $M_L=M_E=M_{\Psi/\Phi}=M$
we obtain the matching relations
\begin{align}
\label{matching}
 C_{e\phi}^\text{I}&=\frac{\lambda_L^\text{I}(\lambda_{E}^\text{I})^* A}{384\pi^2 M^3}\Big(\frac{2|A|^2}{M^2}\xi_{e\phi}+\lambda_H\,\xi_{eB}\Big),\notag\\
 C_{e\phi}^\text{II}&=-\frac{\lambda_L^\text{II}(\lambda_{E}^\text{II})^*\kappa}{192\pi^2 M^2}\Big(2|\kappa|^2\,\xi_{e\phi}+\lambda_H\,\xi_{eB}\Big),\notag\\
C_{eB}^\text{I}&=\frac{\lambda_L^\text{I}(\lambda_{E}^\text{I})^*A g'}{768\pi^2 M^3}\xi_{eB}\Big(Y_L+Y_E-2Y_\Psi\Big),\notag\\
C_{eB}^\text{II}&=-\frac{\lambda_L^\text{II}(\lambda_{E}^\text{II})^*\kappa g'}{384\pi^2 M^2}\xi_{eB}\Big(Y_L
+Y_E\Big),\notag\\
C_{eW}^\text{I}&=\frac{\lambda_L^\text{I}(\lambda_{E}^\text{I})^*A g}{768\pi^2 M^3}\Big(\xi_{eW}^L+\xi_{eW}^E-2\tilde \xi_{eW}\Big),\notag\\
C_{eW}^\text{II}&=-\frac{\lambda_L^\text{II}(\lambda_{E}^\text{II})^*\kappa g}{384\pi^2 M^2}\Big(\xi_{eW}^L
+\xi_{eW}^E\Big),\notag\\
C_{\phi e}^{\text{I}}&=\frac{|\lambda_E^\text{I}|^2}{768\pi^2 M^2}\bigg[\frac{|A|^2}{M^2}\xi_{\phi e}
+2\xi_{\phi e}'\,g'^2Y_\phi\Big(Y_{E}+3Y_{\Psi}\Big)\bigg],\notag\\
C_{\phi e}^{\text{II}}&=\frac{|\lambda_E^\text{II}|^2|}{384\pi^2 M^2}\bigg[4|\kappa|^2\,\xi_{\phi e}+\xi_{\phi e}'\,g'^2Y_\phi
\Big(3Y_{E}+Y_{\Phi}\Big)\bigg],\notag\\
C_{\phi \ell}^{(1)\text{I}}&=\frac{|\lambda_L^\text{I}|^2}{768\pi^2 M^2}\bigg[\frac{|A|^2}{M^2}\xi_{\phi\ell}^{(1)}+2\xi_{\phi\ell}'^{(1)}\,g'^2Y_\phi \Big(Y_{L}+3Y_{\Psi}\Big)\bigg],\notag\\
C_{\phi \ell}^{(1)\text{II}}&=\frac{|\lambda_L^\text{II}|^2}{384\pi^2 M^2}\bigg[4|\kappa|^2\,\xi_{\phi\ell}^{(1)}
+\xi_{\phi\ell}'^{(1)}\,g'^2Y_\phi\Big(3Y_{L}+Y_{\Phi}\Big)\bigg],\notag\\
C_{\phi \ell}^{(3)\text{I}}&=\frac{|\lambda_L^\text{I}|^2}{768\pi^2 M^2}\bigg[\frac{|A|^2}{M^2}\xi_{\phi\ell}^{(3)}+2g^2
\Big(\xi_{\phi\ell}'^{(3)} +3\xi_{\phi\ell}''^{(3)}\Big)\bigg],\notag\\
C_{\phi \ell}^{(3)\text{II}}&=\frac{|\lambda_L^\text{II}|^2}{384\pi^2 M^2}\bigg[4|\kappa|^2\,\xi_{\phi\ell}^{(3)}
+g^2\Big(3\xi_{\phi\ell}'^{(3)} +\xi_{\phi\ell}''^{(3)}\Big)\bigg],
\end{align}
while the result for general masses is given in Eq.~\eqref{general_matching}.  The representation-dependent $SU(2)_L$ factors $\xi$ are collected in Table~\ref{xi_R}, and the quartic Higgs couplings is defined by $\lambda_H=M_H^2/v^2$ (with the vev normalization
$v\sim 246\GeV$). In terms of these coefficients the modifications to the observables are as given in App.~\ref{sec:master}. Note that we performed an on-shell matching. This means that after EW symmetry breaking the $\lambda_H$ terms in the first two Wilson coefficients and the $g^2$ and $g'^2$ terms in the last six ones only contribute to physical $W$, $Z$, and Higgs decays but are absent in effective couplings at zero momentum transfer.
In principle one could now apply the renormalization group evolution within SMEFT~\cite{Jenkins:2013zja,Alonso:2013hga,Aebischer:2021uvt} down to the EW scale. However, as we assume TeV-scale new physics, this effect turns out to be small. Furthermore, even the evolution of the magnetic operator from the EW scale to the muon scale only amounts to a reduction of $\approx 6\%$~\cite{Crivellin:2017rmk}, which we include in our analysis.

\begin{figure*}[t]
	\includegraphics[width=0.5\linewidth]{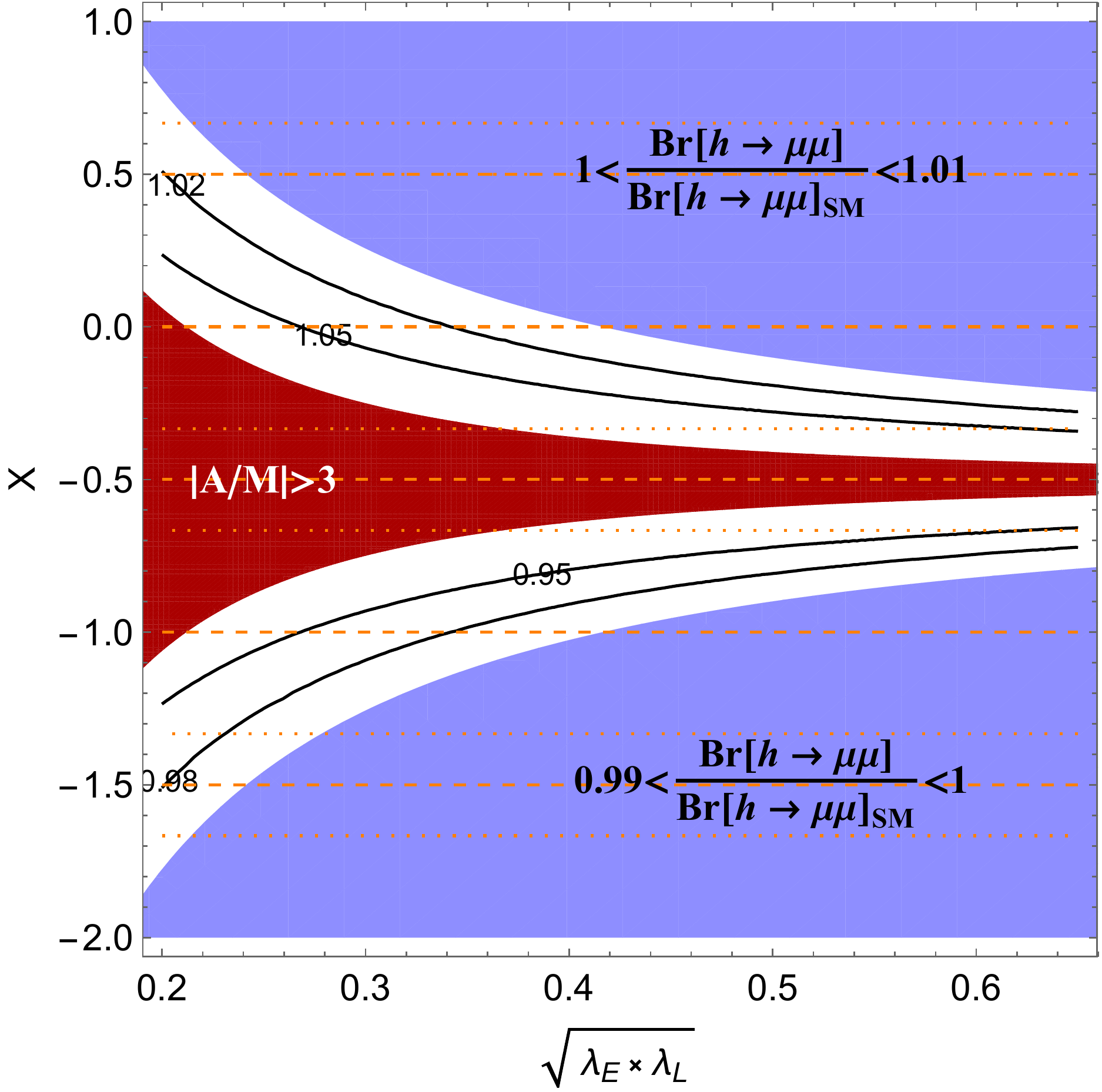}
	~~
	\includegraphics[width=0.475\linewidth]{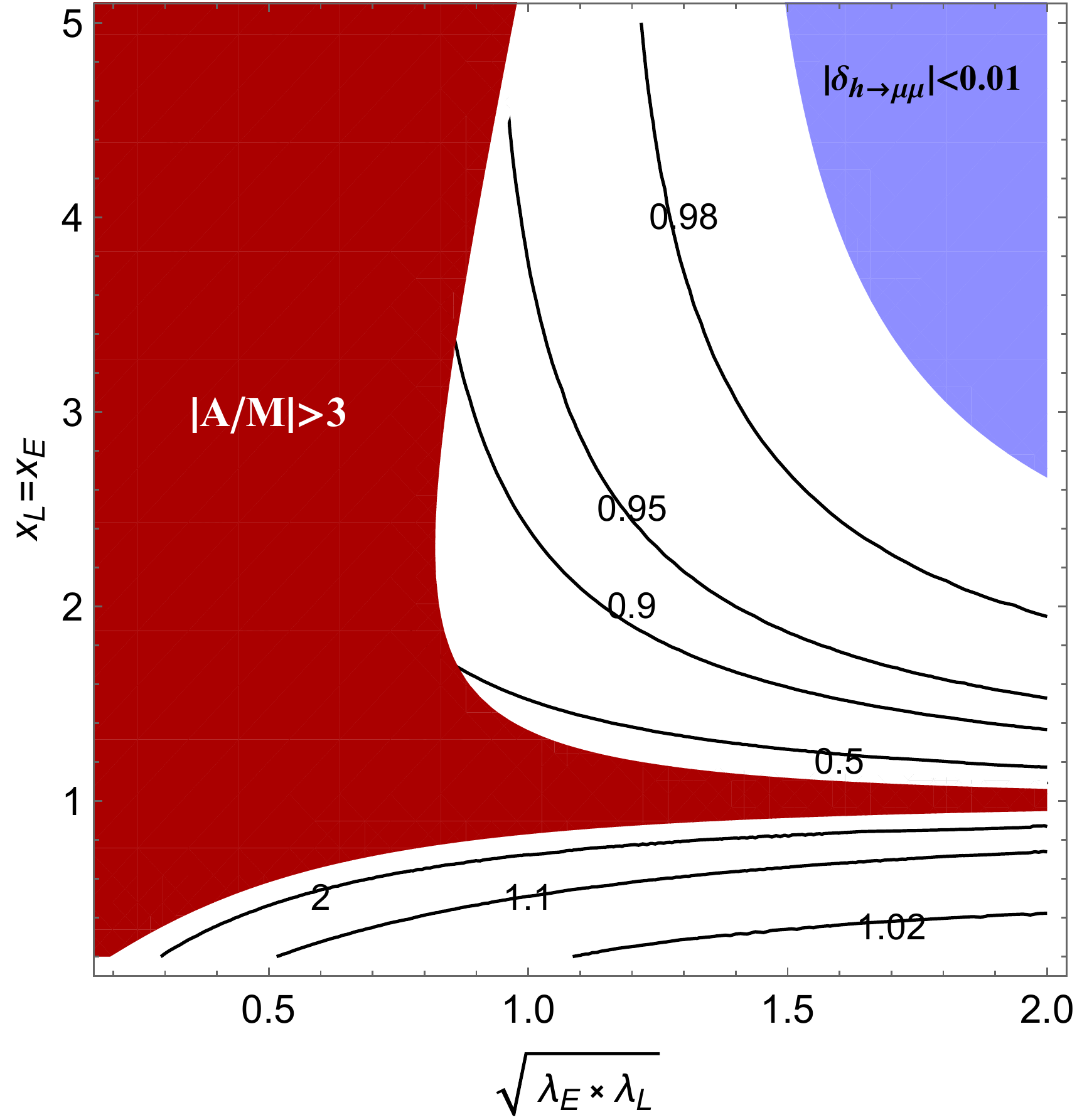}
	\caption{Modification of $h\to\mu\mu$ for $R=121$ in case I, as a function of $\sqrt{\lambda_L^\text{I}\lambda_E^\text{I}}$ and the hypercharge $X$ (left), and  for fixed $X=-1/2$, but with variable mass parameters $x_{L,E}=M_{L,E}^2/M_\Psi^2$ (right). In both cases regions in parameter space where the modification stays below $1\%$ are marked in blue, those with coupling $|A/M|>3$ in red. The mass parameter is set to $M=1\TeV$, and the central value of $\Delta a_\mu$ has been assumed.} 
	\label{fig:hmumu}
\end{figure*}

\section{Phenomenology and Correlations}
\label{sec:correlations}

After EW symmetry breaking the operators and the associated Wilson coefficients derived in the last section contribute to the observables $(g-2)_\mu$, $Z\to\mu\mu$, and $h\to\mu\mu$.

\subsection[$(g-2)_\mu$]{$\boldsymbol{(g-2)_\mu}$}

The contribution to $(g-2)_\mu$ can be written as
\begin{align}
 \Delta a_\mu^\text{I}&=-\frac{m_\mu v\,\Re[\lambda_L^\text{I}(\lambda_E^\text{I})^* A]}{96\pi^2\sqrt{2}\,M^3} 
 \Big\{2X+1,-2X,6X-1,2(3X+2)\Big\},\notag\\ 
 \Delta a_\mu^\text{II}&=\frac{m_\mu v\,\Re[\lambda_L^\text{II}(\lambda_E^\text{II})^* \kappa]}{96\pi^2\sqrt{2}\,M^2} \Big\{2(X+1),-(2X+1),2(3X+1),6X+7\Big\},
 \label{Deltaamu}
\end{align}
for case I and II, respectively, where the brackets give the hypercharge factors for the representations $R$ (in the same order as in Table~\ref{xi_R}). For some hypercharges  $\Delta a_\mu$ thus vanishes in the equal-mass limit, in which case the full expressions from Eq.~\eqref{general_matching} show that the cancellation is lifted by mass effects.
The chiral enhancement is reflected by the fact that $v$ enters multiplied by a free parameter (possibly of order one) not related to the muon Yukawa coupling. We stress that the phase of the coupling is not determined, and while we will concentrate on the real part in this paper, which interferes with the SM contribution, it is this phase $\delta$ that would generate a muon electric dipole moment (EDM) $d_\mu$~\cite{Crivellin:2018qmi}, see also Refs.~\cite{Dekens:2018bci,Crivellin:2019qnh,Altmannshofer:2020ywf,Babu:2020hun}. For $\tan\delta=\Order(1)$, the resulting $|d_\mu|\sim 10^{-22}e\,\text{cm}$ could be measured with the  proposed  muon  EDM  experiment at PSI using the frozen-spin technique~\cite{Adelmann:2021udj}.

\subsection[Correlations with $h\to \mu\mu$]{Correlations with $\boldsymbol{h\to \mu\mu}$}

\begin{figure*}[t]
	\includegraphics[width=0.48\linewidth]{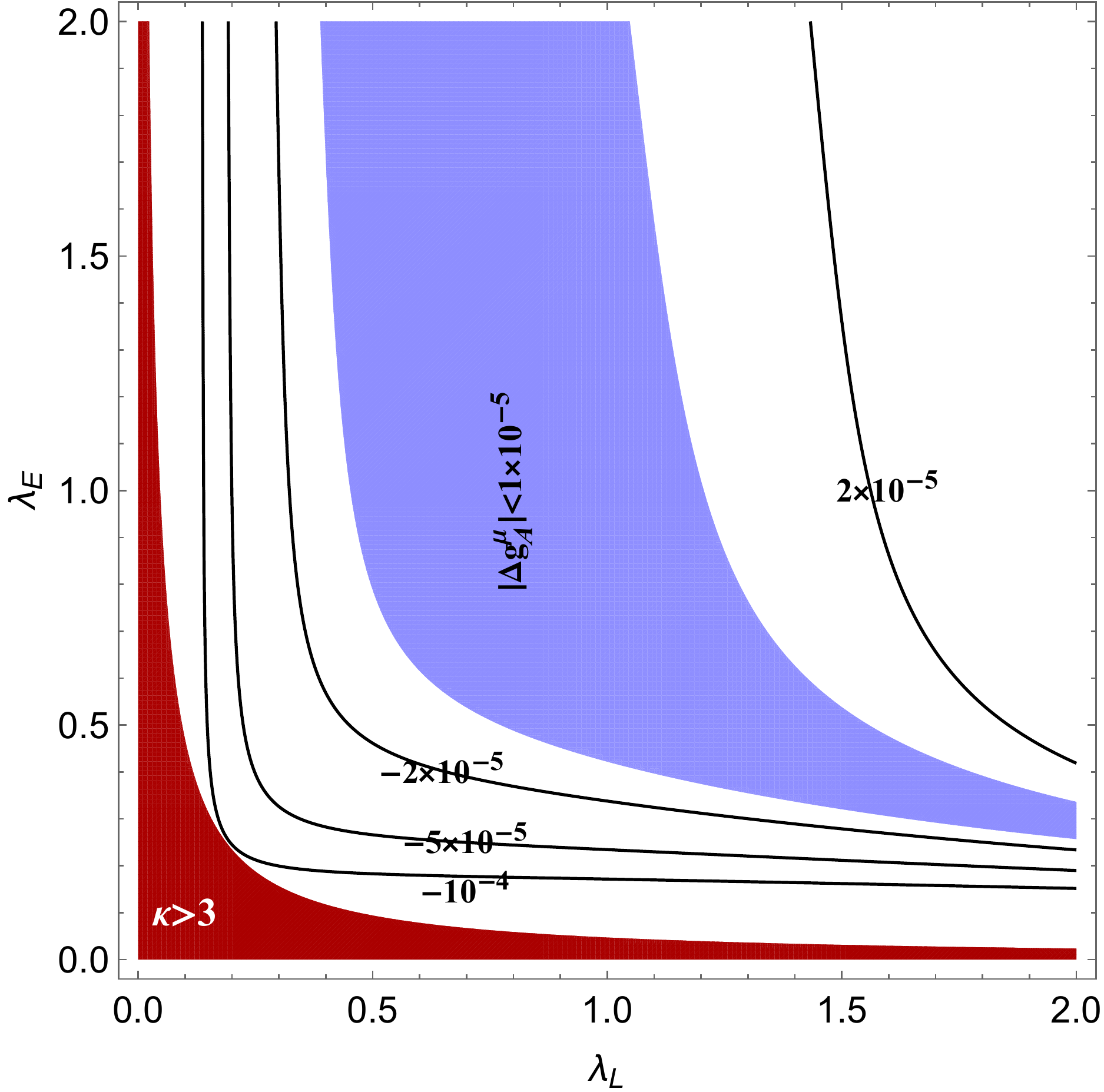}~~~
	\includegraphics[width=0.48\linewidth]{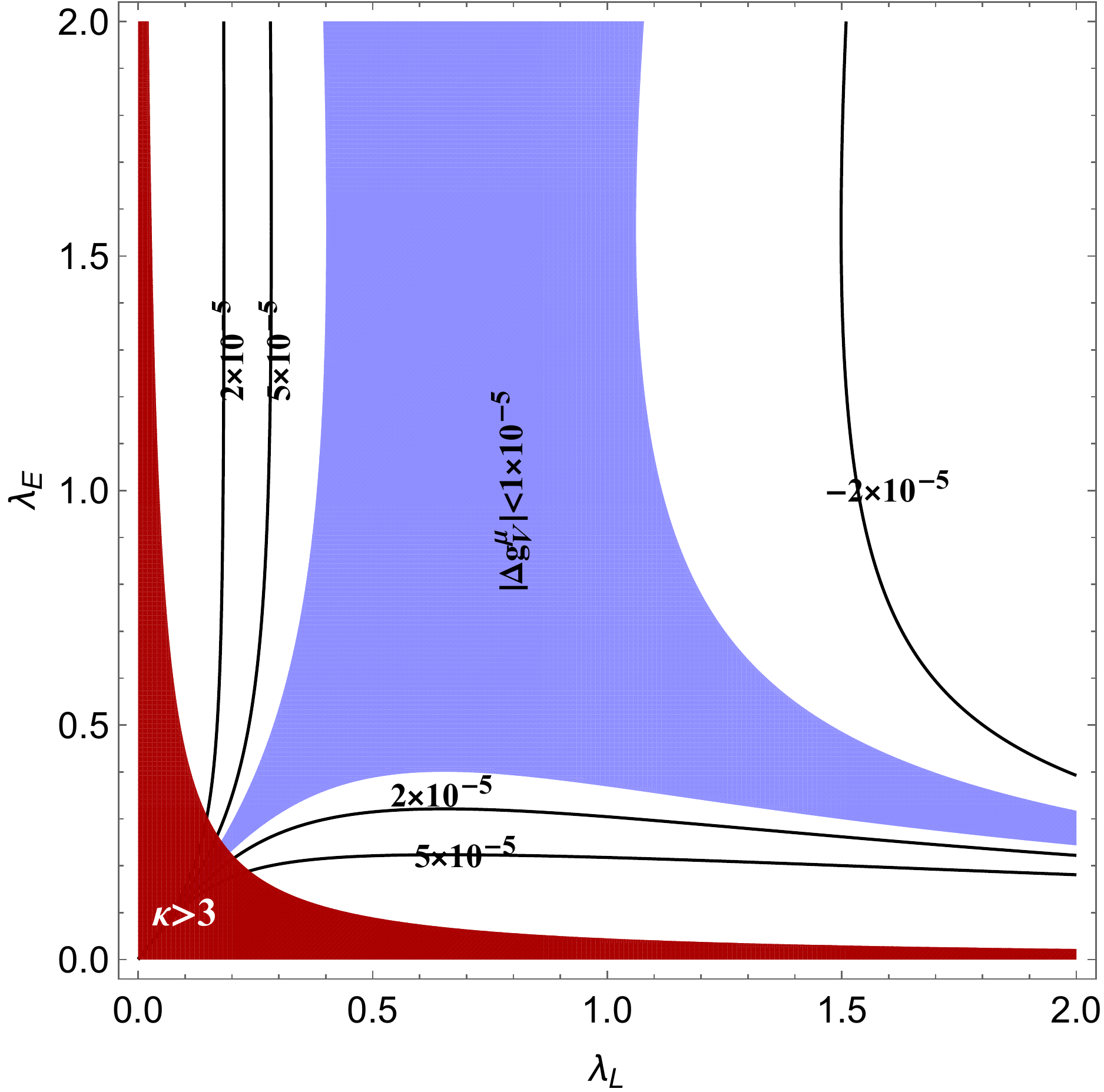}
	\caption{Modification of $g_A^\mu$ (left) and $g_V^\mu$ (right) for $R=232$ in case II, as a function of $\lambda_E^\text{II}$ and $\lambda_L^\text{II}$ for fixed hypercharge $X=0$. 
	 In both cases regions in parameter space where the modification stays below $10^{-5}$ are marked in blue, those with coupling $|\kappa|>3$ in red.} 
	\label{fig:Zmumu}
\end{figure*}

We express the modification to the $h\to\mu\mu$ decay rate as
\begin{equation}
 \frac{\Br[h\to\mu\mu]}{\Br[h\to\mu\mu]|_\text{SM}}\equiv1+\delta_{h\to\mu\mu}=1-\frac{\sqrt{2}\, C_{e\phi} v^3}{m_\mu}.
\end{equation}
With $\xi_{eB}$ and $\xi_{e\phi}$  having the same sign in all cases, it then follows from Eq.~\eqref{matching} that the effect in $h\to\mu\mu$ cannot be canceled between the two terms (for equal masses). In fact, in case I the effect for $R=121,323,232$ ($R=212$) is necessarily destructive (constructive) and vice versa for case II. Furthermore, the contribution originating from $\lambda_H$ is completely determined by the effect in $\Delta a_\mu$. However, it turns out to be numerically small, in general below $1\%$, and thus beyond the reach of future colliders (with current measurements of the branching fraction relative to the SM at $1.2(6)$~\cite{Aad:2020xfq} and $1.2(4)$
\cite{Sirunyan:2020two}).  

Nonetheless, larger effects can occur when both terms are included, but then a dependence on the couplings, shown in Fig.~\ref{fig:hmumu} as a function of $\lambda^\text{I}=\sqrt{\lambda_L^\text{I}\lambda_E^\text{I}}$ (with $A/M$ determined from the central value of $\Delta a_\mu$), arises. For equal masses the biggest modification occurs for small $\lambda^\text{I}$ in the vicinity of the hypercharge for which $\Delta a_\mu$ vanishes, and the effects for the representations $R=121,212$ in general exceed those for $R=323,232$. We also show an example for a case in which the mass difference between $M_{L,E}$ and $M_\Psi$ lifts the cancellation in $\Delta a_\mu$, which can also lead to sizable effects within reach of future $e^+e^-$ colliders~\cite{Aicheler:2012bya,Baer:2013cma,An:2018dwb,Abada:2019zxq}, while percent-level precision would require the FCC-hh~\cite{Benedikt:2018csr}. While the guaranteed effect in $h\to\mu\mu$ is thus small, there are regions in parameter space in which a deviation from the SM could be detected, and thus allow one to distinguish between different scenarios.   

\subsection[Correlations with $Z\to \mu\mu$]{Correlations with $\boldsymbol{Z\to \mu\mu}$}

For $Z\to\mu\mu$ the correlation with $(g-2)_\mu$ is less direct, as the left- and right-handed couplings $\lambda_{E,L}$ appear separately in the contributions from $C_{\phi e}$ and $C_{\phi\ell}^{(1),(3)}$, see Eq.~\eqref{gVgA}.\footnote{Note that the modification of the off-shell $W$--$\mu$--$\nu$ coupling implied by $SU(2)_L$ symmetry also leads to a shift in the Fermi constant determined from muon decay, which enters the global EW fit. Even though this effect is small, the modifications of $g_{V,A}^\mu$ should therefore be understood as a relative shift with respect to $g_{V,A}^e$.}  
With the product $\Re[\lambda_L^\text{II} (\lambda_E^\text{II})^* \kappa]$ (or $\Re[\lambda_L^\text{I}( \lambda_E^\text{I})^* A]$) determined from $\Delta a_\mu$, the relative size of $\lambda_{E,L}$ thus matters. In particular, if there is a relative sign between the left- and right-handed contributions, with coefficients of similar size, the effect in $Z\to\mu\mu$ is minimized for $\lambda_E\sim \lambda_L$, an example for which is shown in Fig.~\ref{fig:Zmumu} (for $R=232$ in case II with $X=0$). Here, the vector coupling $g_V^\mu$ displays this behavior for small $\lambda_{E,L}^\text{II}$ before the interplay with the $|\kappa|^2$ term becomes relevant, while for the axial-vector coupling $g_A^\mu$ both contributions enter with the same sign. Overall, there are regions in parameter space in which the effect is suppressed below the level of $10^{-5}$, but we do find large areas that can be probed by future colliders. In general, we find that the effects are larger for $R=323,232$ than for $R=121,212$, contrary to $h\to\mu\mu$.  

The current limits for the $Z\to\mu\mu$ couplings are~\cite{Zyla:2020zbs,ALEPH:2005ab}
\begin{align}
 \Delta g_A^\mu&=-0.1(5.4)\times 10^{-4}& \big[\Delta g_A^\ell&=-0.4(5.6)\times 10^{-4}\big],\notag\\
 \Delta g_V^\mu&=3.2(23.0)\times 10^{-4}& \big[\Delta g_V^\ell&=-8.1(8.8)\times 10^{-4}\big],
\end{align}
where in brackets we show for comparison the precision when flavor-universal couplings are assumed. Neither constraint is visible in Fig.~\ref{fig:Zmumu}, which emphasizes the importance of future experiments, promising improvements by two orders of magnitude~\cite{Aicheler:2012bya,Baer:2013cma,An:2018dwb,Abada:2019zxq}, to constrain the relevant parameter space.

\subsection[Implications from $SU(2)_L$ symmetry]{Implications from $\boldsymbol{SU(2)_L}$ symmetry}

The operators $C_{\phi\ell}^{(1),(3)}$ also induce modifications in $Z\to\nu\nu$ and $W\to\ell\nu$, with the latter solely generated by $C_{\phi\ell}^{(3)}$. 
In the case of $Z\to\nu\nu$ the current constraints are~\cite{Zyla:2020zbs,Vilain:1993xb,ALEPH:2005ab}
\begin{align}
 \Delta g^{\nu_\mu}=0.2(1.7)\times 10^{-2}\; [\Delta g^{\nu_\ell}=7.6(7.6)\times 10^{-4}],
\end{align}
where the number in brackets gives the flavor-universal limit. Especially when concentrating on $\nu_\mu$, it is therefore clear that the $Z\to\nu\nu$ channel is less constraining than $Z\to\mu\mu$. 

In view of the typical size of the corrections found for $h\to\mu\mu$ and $Z\to\mu\mu$, the apparent violation of CKM unitarity~\cite{Belfatto:2019swo,Grossman:2019bzp,Crivellin:2020lzu,Crivellin:2021njn} cannot be explained by our modified off-shell $W$--$\mu$--$\nu$ couplings, for which significantly larger effects are needed~\cite{Coutinho:2019aiy,Endo:2020tkb,Crivellin:2020ebi,Kirk:2020wdk,Alok:2020jod,Crivellin:2021egp}. In particular, the only remaining contribution to $C_{\phi \ell}^{(3)}$, the one involving $|\kappa|^2$ or $|A|^2$, either vanishes or is negative, due to the coefficient $\xi_{\phi\ell}^{(3)}$, while alleviating the CKM-unitarity tension requires $\varepsilon_{\mu\mu}>0$.

\section{Conclusions and Outlook}
\label{sec:conclusions}

In this article we discussed explanations of the anomalous magnetic moment of the muon in terms of BSM physics with heavy, TeV-scale new scalars and fermions. As the deviation from the SM prediction of $\Delta a_\mu=251(59)\times 10^{-11}$ is of the order of the EW contribution, an enhancement factor is necessary to account for it. Such a mechanism can be provided if the BSM theory involves couplings to the SM Higgs that are much larger than the muon Yukawa coupling. For this chiral enhancement, in a generic model with new scalars and fermions, at least three representations under $SU(2)_L\times U(1)_Y$, with two new scalars and one new fermion (case I) or two new fermions and one new scalar (case II), are necessary, see Table~\ref{quantum_numbers}. In addition, in order to avoid prohibitively large tree-level effects in $Z$ or Higgs decays, we assumed that our model possesses an additional $Z_2$ symmetry. For the hypercharges
 listed in Table~\ref{dark_matter} 
electrically neutral states exist among the various components of the $SU(2)_L$ representations, which could at the same time provide viable dark matter candidates.

Within this setup we performed the matching onto SMEFT (within unbroken $SU(2)_L$) for all operators involving two muons, which even for general hypercharges, masses, and different $SU(2)_L$ representations can be written in a surprisingly compact form.   
In addition to the loop functions, which reduce to simple numerical factors in the limit of equal masses, the resulting Wilson coefficients, see Eq.~\eqref{matching}, differ between cases I and II, as well as factors resulting from the $SU(2)_L$ representations, see Table~\ref{xi_R}, covering a wide range of BSM scenarios that display chiral enhancement in $(g-2)_\mu$.  

\begin{table}[t!]
	\centering
	\begin{tabular}{ccccc}
		\toprule
		$R$& $121$ & $212$ & $323$ & $232$\\
		$X$& $\big\{-1,0\big\}$
		&$\big\{-\frac{3}{2},-\frac{1}{2},\frac{1}{2}\big\}$
		&$\big\{-2,-1,0,1\big\}$&$\big\{-\frac{3}{2}, -\frac{1}{2},\frac{1}{2}\big\}$\\
		\bottomrule
	\end{tabular}
	\caption{Possible hypercharges for each representation that could lead to a dark matter candidate.}
	\label{dark_matter}
\end{table}

The Wilson coefficients are then related to $(g-2)_\mu$, $Z\to\mu\mu$, $h\to\mu\mu$, and give rise to modified $W$--$\mu$--$\nu$ couplings, as we studied in our phenomenological analysis.  Imposing $\Delta a_\mu$ on the parameters, we find that (for equal masses) $R=121$, $323$, $232$ ($R=121$, $323$, $232$) predict a constructive (destructive) effect in $h\to\mu \mu$ for case I and vice versa for case II, see Fig.~\ref{fig:hmumu} for an exemplary case. For $Z\to\mu\mu$ the correlations in addition depend on the ratio of left- and right-handed couplings, see Fig.~\ref{fig:Zmumu} for the two typical patterns that emerge depending on the relative sign of the left- and right-handed contributions. In both $h\to\mu\mu$ and $Z\to\mu\mu$ the effects lie below the current experimental precision, in line with the inherent loop suppression, but depending on $SU(2)_L$ representation, hypercharge, and couplings  
 are expected to be at a level observable by a future $e^+e^-$ collider~\cite{Aicheler:2012bya,Baer:2013cma,An:2018dwb,Abada:2019zxq} and the FCC-hh~\cite{Benedikt:2018csr} in $Z\to\mu\mu$ and $h\to\mu\mu$, respectively. The interplay with precision flavor physics at future colliders would thus allow one to derive valuable hints regarding the nature of BSM physics that may be hidden in $(g-2)_\mu$ and can be used to enhance the physics case for such machines.

\begin{acknowledgments}
We thank Athanasios Dedes and Christoph Greub for useful discussions, and Matthew Kirk for checking our results using the software \texttt{Matchmakereft}~\cite{Carmona:2021xtq}. Support by the Swiss National Science Foundation, under Project Nos.\ PP00P21\_76884 (A.C.) and PCEFP2\_181117 (M.H.), is gratefully acknowledged.
\end{acknowledgments}

\appendix

\section{Master formulae}
\label{sec:master}

The Higgs decay rate is given by $C_{e\phi}$
\begin{equation}
 \Gamma[h\to\mu\mu]=\frac{M_H}{8\pi}\sqrt{1-\frac{4m_\mu^2}{M_H^2}}\bigg(\frac{m_\mu}{v}-\frac{C_{e\phi} v^2}{\sqrt{2}}\bigg)^2,
\end{equation}
corresponding to a relative change in the Higgs Yukawa coupling of 
\begin{equation}
 \frac{\delta Y_\mu}{Y_\mu}=-C_{e\phi}\frac{v^3}{\sqrt{2}\,m_\mu},
\end{equation}
and the contribution to the anomalous magnetic moment reads
\begin{equation}
 \Delta a_\mu=\frac{4m_\mu v}{\sqrt{2}\, e}\Big(\Re \big[C_{eB}\big] c_W-\Re \big[C_{eW}\big] s_W\Big),
\end{equation}
where $c_W=\cos\theta_W$, $s_W=\sin\theta_W$.
The modifications to the $Z$ and $W$ couplings are
\begin{align}
 {\mathcal L}_Z&=\frac{g}{2c_W}\bigg[1-2s_W^2+v^2\Big(C_{\phi\ell}^{(1)}+C_{\phi\ell}^{(3)}\Big)\bigg]\bar\mu \gamma^\mu P_L \mu \,Z_\mu\notag\\
 &+\frac{g}{2c_W}\Big[-2s_W^2+v^2 C_{\phi e}\Big]\bar \mu \gamma^\mu P_R \mu\, Z_\mu\notag\\
 &-\frac{g}{2c_W}\bigg[1-v^2\Big(C_{\phi\ell}^{(1)}-C_{\phi\ell}^{(3)}\Big)\bigg]\bar\nu\gamma^\mu P_L\nu\, Z_\mu,\notag\\
  {\mathcal L}_W&=-\frac{g}{\sqrt{2}}\Big(1+v^2 C_{\phi\ell}^{(3)}\Big)\Big(W_\mu^- \, \bar\mu \gamma^\mu P_L \nu + W_\mu^+ \, \bar\nu \gamma^\mu P_L \mu\Big).
\end{align}
Writing the interactions as~\cite{Zyla:2020zbs} 
\begin{align}
{\mathcal L}_Z&=-\frac{g}{2c_W}
\bar f\gamma^\mu \big(g_V^f - g_A^f \gamma_5\big)f\,Z_\mu,\notag\\
{\mathcal L}_W&=-\frac{g}{\sqrt{2}}\big(1+\varepsilon_{\mu\mu}\big)\Big(W_\mu^- \, \bar\mu \gamma^\mu P_L \nu + W_\mu^+ \, \bar\nu \gamma^\mu P_L \mu\Big),
\end{align}
we have
\begin{align}
\label{gVgA}
 g_V^\mu&=-\frac{1}{2}\big(1-4s_W^2\big)-\frac{v^2}{2}\Big(C_{\phi\ell}^{(1)}+C_{\phi\ell}^{(3)}+C_{\phi e}\Big),\notag\\
 g_A^\mu&=-\frac{1}{2}-\frac{v^2}{2}\Big(C_{\phi\ell}^{(1)}+C_{\phi\ell}^{(3)}-C_{\phi e}\Big),\notag\\
 g^\nu&\equiv g_V^\nu=g_A^\nu=\frac{1}{2}-\frac{v^2}{2}\Big(C_{\phi\ell}^{(1)}-C_{\phi\ell}^{(3)}\Big),\notag\\ 
 \varepsilon_{\mu\mu}&=v^2 C_{\phi\ell}^{(3)}.
\end{align} 

\section{Simplified models and matching relations}
\label{sec:simp}

The different $SU(2)_L$ representations are implemented according to
\begin{align}
\Lagr_\text{I}^{121} &= \lambda _L^\text{I}\,\bar\ell^a\Psi\Phi _L^a + \lambda _E^\text{I}\,\bar e\Psi\Phi _E + A\,\Phi _L^{a\dag }\Phi _E{\phi^a},\notag\\
\Lagr_\text{II}^{121} &= \lambda _L^\text{II}\,\bar\ell^a\Phi\Psi _L^a + \lambda _E^\text{II}\,\bar e\Psi _E{\Phi } + \kappa\, \bar \Psi _L^a\Psi _E{\phi^a},\notag\\
\Lagr_\text{I}^{212} &= \lambda _L^\text{I}\,\bar\ell^a\Phi _L{\Psi ^a} + \lambda _E^\text{I}\,\bar e\Psi _{}^a(i\tau_2\Phi _E)^a + A\,{(i\tau_2 \phi)^a} \Phi _L^\dag \Phi _E^{a},\notag\\
\Lagr_\text{II}^{212} &= \lambda _L^\text{II}\,\bar\ell^a\Psi _L{\Phi ^a} + \lambda _E^\text{II}\,\bar e(i\tau_2\Psi _E)^a{\Phi^a} + \kappa\, \bar \Psi_L{\left( {i{\tau _2}\phi} \right)^a}\Psi _E^a,\notag\\
\Lagr_\text{I}^{323} &= \lambda _L^\text{I}\,\bar\ell^a{\left( {\tau \cdot\Psi } \right)_{ab}}\Phi _L^b + \lambda _E^\text{I}\,\bar e\Psi _{}^\alpha\Phi _E^\alpha + A\,\Phi _L^{a\dag }{\left( {\tau \cdot\Phi _E^{}} \right)_{ab}}{\phi^b},\notag\\
\Lagr_\text{II}^{323} &= \lambda _L^\text{II}\,\bar\ell^a{\left( {\tau \cdot\Phi } \right)_{ab}}\Psi _L^b + \lambda _E^\text{II}\,\bar e\Psi _E^\alpha{\Phi ^\alpha} + \kappa\, \bar \Psi _L^a{\left( {\tau \cdot{\Psi _E}} \right)_{ab}}{\phi^b},\notag\\
\Lagr_\text{I}^{232} &= \lambda _L^\text{I}\,\bar\ell^a{\left( {\tau \cdot{\Phi _L}} \right)_{ab}}{\Psi ^b} + \lambda _E^\text{I}\,\bar e\Psi _{}^a(i\tau_2\Phi _E)^a + A\,(i\tau_2 \phi)^a(\tau\cdot \Phi _L^\dagger)_{ab}\Phi _E^{b},\notag\\
\Lagr_\text{II}^{232} &= \lambda _L^\text{II}\,\bar\ell^a{\left( {\tau \cdot\Psi _L^{}} \right)_{ab}}{\Phi ^b} + \lambda _E^\text{II}\,\bar e(i\tau_2\Psi _E)^a{\Phi^a} + \kappa\, {\left( {\tau \cdot{{\bar \Psi }_L}} \right)_{ab}}{\left( {i{\tau _2}\phi} \right)^a}\Psi _E^b,
\label{SU2contractions}
\end{align}
where $\tau$ denotes the Pauli matrices.
The matching relations for general masses read 
\begin{align}
C_{e\phi}^\text{I}&=\frac{\lambda_L^\text{I}(\lambda_{E}^\text{I})^* A}{16\pi^2 M_{\Psi}^3}\bigg[\frac{|A|^2}{M_\Psi^2}\xi_{e\phi}\,f_{e\phi}(x_L,x_E) 
+ \lambda_H\xi_{eB}\,\tilde f_{e\phi}(x_L,x_E)\bigg],\notag\\
 C_{e\phi}^\text{II}&=\frac{\lambda_L^\text{II}(\lambda_{E}^\text{II})^*\kappa}{16\pi^2 M_{\Phi}^2}\bigg[|\kappa|^2\xi_{e\phi}\,g_{e\phi}
(x_L,x_E)
+\lambda_H\xi_{eB}\,\tilde g_{e\phi}
(x_L,x_E)\bigg],\notag\\
C_{eB}^\text{I}&=\frac{\lambda_L^\text{I}(\lambda_{E}^\text{I})^*A g'}{16\pi^2 M_{\Psi}^3}\xi_{eB}\bigg[Y_L\, f_{eB}(x_L,x_E)+Y_E\, f_{eB}(x_E,x_L)
+Y_\Psi\, \tilde f_{eB}(x_L,x_E)\bigg],\notag\\
C_{eB}^\text{II}&=\frac{\lambda_L^\text{II}(\lambda_{E}^\text{II})^*\kappa g'}{16\pi^2 M_{\Phi}^2}\xi_{eB}\bigg[Y_L\,g_{eB}(x_L,x_E)
+Y_E\,g_{eB}(x_E,x_L)
+Y_{\Phi}\,\tilde g_{eB}(x_L,x_E)\bigg],\notag\\
C_{eW}^\text{I}&=\frac{\lambda_L^\text{I}(\lambda_{E}^\text{I})^*A g}{16\pi^2 M_{\Psi}^3}\bigg[\xi_{eW}^L\,f_{eB}(x_L,x_E)+\xi_{eW}^E\,f_{eB}(x_E,x_L)
+\tilde \xi_{eW}\,\tilde f_{eB}(x_L,x_E)\bigg],\notag\\
C_{eW}^\text{II}&=\frac{\lambda_L^\text{II}(\lambda_{E}^\text{II})^*\kappa g}{16\pi^2 M_{\Phi}^2}\bigg[\xi_{eW}^L\,g_{eB}(x_L,x_E)
+\xi_{eW}^E\,g_{eB}(x_E,x_L)+\tilde \xi_{eW}\,\tilde g_{eB}(x_L,x_E)\bigg],\notag\\
C_{\phi e}^{\text{I}}&=\frac{|\lambda_E^\text{I}|^2}{16\pi^2 M_{\Psi}^2}\bigg[\frac{|A|^2}{M_\Psi^2}\xi_{\phi e}
\,f_{\phi e}(x_L,x_E)
+\xi_{\phi e}'\,g'^2Y_\phi\Big(Y_{E} \,h_{\phi e}(x_E)+Y_{\Psi}\tilde h_{\phi e}(x_E)\Big)\bigg],\notag\\
C_{\phi e}^{\text{II}}&=\frac{|\lambda_E^\text{II}|^2}{16\pi^2 M_{\Phi}^2}\bigg[|\kappa|^2\xi_{\phi e}
\,g_{\phi e}(x_L,x_E)
+\xi_{\phi e}'\,g'^2Y_\phi\Big(Y_{E} \,k_{\phi e}(x_E)+Y_{\Phi} \tilde k_{\phi e}(x_E)\Big)\bigg],\notag\\
C_{\phi \ell}^{(1)\text{I}}&=\frac{|\lambda_L^\text{I}|^2}{16\pi^2 M_{\Psi}^2}\bigg[\frac{|A|^2}{M_\Psi^2}\xi_{\phi\ell}^{(1)}\,f_{\phi e}(x_L,x_E)
+\xi_{\phi\ell}'^{(1)}\,g'^2Y_\phi\Big(Y_{L} \,h_{\phi e}(x_L)+Y_{\Psi}\,\tilde h_{\phi e}(x_L)\Big)\bigg],\notag\\
C_{\phi \ell}^{(1)\text{II}}&=\frac{|\lambda_L^\text{II}|^2}{16\pi^2 M_{\Phi}^2}\bigg[|\kappa|^2\xi_{\phi\ell}^{(1)}\,g_{\phi e}(x_L,x_E)
+\xi_{\phi\ell}'^{(1)}\,g'^2Y_\phi\Big(Y_{L} \,k_{\phi e}(x_L)+Y_{\Phi}\,\tilde k_{\phi e}(x_L)\Big)\bigg],\notag\\
C_{\phi \ell}^{(3)\text{I}}&=\frac{|\lambda_L^\text{I}|^2}{16\pi^2 M_{\Psi}^2}\bigg[\frac{|A|^2}{M_\Psi^2}\xi_{\phi\ell}^{(3)}\,f_{\phi e}(x_L,x_E)
+g^2\Big(\xi_{\phi\ell}'^{(3)} \,h_{\phi e}(x_L)+ \xi_{\phi\ell}''^{(3)}\,\tilde h_{\phi e}(x_L)\Big)\bigg],\notag\\
C_{\phi \ell}^{(3)\text{II}}&=\frac{|\lambda_L^\text{II}|^2}{16\pi^2 M_{\Phi}^2}\bigg[|\kappa|^2\xi_{\phi\ell}^{(3)}\,g_{\phi e}(x_L,x_E)
+g^2\Big(\xi_{\phi\ell}'^{(3)} \,k_{\phi e}(x_L)+\xi_{\phi\ell}''^{(3)}\,\tilde k_{\phi e}(x_L)\Big)\bigg],
\label{general_matching}
\end{align} 
with representation-dependent factors $\xi$ given in Table~\ref{xi_R}, 
$x_{L,E}=M_{L,E}^2/M_\Psi^2$ and $x_{L,E}=M_{L,E}^2/M_\Phi^2$ for cases I and II, respectively,
and loop functions
\begin{align}
f_{e\phi}(x,y)&=\frac{x+y-2}{(x-1)(y-1)(x-y)^2}+\frac{(x+y-2x^2)\log x}{(x-1)^2(x-y)^3}-\frac{(x+y-2y^2)\log y}{(y-1)^2(x-y)^3},\notag\\
 g_{e\phi}(x,y)&=-\frac{2(x+y-2x y)}{(x-1)(y-1)(\sqrt{x}-\sqrt{y})^2}
 -\frac{x[\sqrt{y}(x-3)+\sqrt{x}(x+1)]\log x}{(x-1)^2(\sqrt{x}-\sqrt{y})^3}\notag\\
 &+\frac{y[\sqrt{x}(y-3)+\sqrt{y}(y+1)]\log y}{(y-1)^2(\sqrt{x}-\sqrt{y})^3},\notag\\
 \tilde f_{e\phi}(x,y)&=2 f_{\phi e}(x,y)=\frac{1}{2}\bigg[\frac{x+y-2xy}{(x-1)(y-1)(x-y)^2}+\frac{x(x^2+(x-2)y)\log x}{(x-1)^2(x-y)^3}\notag\\
 &-\frac{y(y^2+(y-2)x)\log y}{(y-1)^2(x-y)^3}\bigg],\notag\\
 \tilde g_{e\phi}(x,y)&=\frac{1}{4}\bigg[\frac{1}{(x-1)(y-1)}+\frac{x^2\log x}{(x-1)^2(x-y)}-\frac{y^2\log y}{(y-1)^2(x-y)}\bigg]-g_{\phi e}(x,y)\notag\\
 &=\frac{1}{2}\bigg[\frac{x+y-x y-\sqrt{xy}}{(x-1)(y-1)(\sqrt{x}-\sqrt{y})^2}+\frac{x(x+(x-2)\sqrt{xy})\log x}{(x-1)^2(x-y)(\sqrt{x}-\sqrt{y})^2}\notag\\
 &-\frac{y(y+(y-2)\sqrt{xy})\log y}{(y-1)^2(x-y)(\sqrt{x}-\sqrt{y})^2}\bigg],\notag\\
 f_{eB}(x,y)&=-\frac{1}{4}\bigg[\frac{x(y-2)+y}{(x-1)^2(y-1)(x-y)}-\frac{x(x-2y+x^2)\log x}{(x-1)^3(x-y)^2}+\frac{y^2\log y}{(y-1)^2(x-y)^2}\bigg],\notag\\
 g_{eB}(x,y)&=\frac{1}{4}\bigg[\frac{(x y-3y+2)\sqrt{x y}-x(y-2)-y}{(x-1)^2(y-1)(x-y)}+\frac{y^{3/2}(\sqrt{x}(y-2)-\sqrt{y})\log y}{(y-1)^2(x-y)^2}\notag\\
 &+\frac{\sqrt{x}[\sqrt{y}(3x^2-x^3-2y)+\sqrt{x}(x^2+x-2y)]\log x}{(x-1)^3(x-y)^2}
 \bigg],\notag\\
 \tilde f_{eB}(x,y)&=\frac{1}{2}\bigg[\frac{3xy-x-y-1}{2(x-1)^2(y-1)^2}+\frac{x^2\log x}{(x-1)^3(x-y)}-\frac{y^2\log y}{(y-1)^3(x-y)}\bigg],\notag\\
 \tilde g_{eB}(x,y)&=\frac{1}{2}\bigg[
 \frac{(x y+x+y-3)\sqrt{x y}+3x y-x-y-1}{2(x-1)^2(y-1)^2}+\frac{x^{3/2}\log x}{(x-1)^3(\sqrt{x}-\sqrt{y})}\notag\\
 &-\frac{y^{3/2}\log y}{(y-1)^3(\sqrt{x}-\sqrt{y})}\bigg],\notag\\
 f_{\phi e}(x,y)&=\frac{1}{4}\bigg[\frac{x+y-2xy}{(x-1)(y-1)(x-y)^2}+\frac{x(x^2+(x-2)y)\log x}{(x-1)^2(x-y)^3}-\frac{y(y^2+(y-2)x)\log y}{(y-1)^2(x-y)^3}\bigg],\notag\\
 g_{\phi e}(x,y)&=\frac{1}{4}\bigg[\frac{2xy-x-y}{(x-1)(y-1)(\sqrt{x}-\sqrt{y})^2}+\frac{x[x(x+y-2)-4(x-1)\sqrt{xy}]\log x}{(x-1)^2(x-y)(\sqrt{x}-\sqrt{y})^2}\notag\\
 &-\frac{y[y(x+y-2)-4(y-1)\sqrt{xy}]\log y}{(y-1)^2(x-y)(\sqrt{x}-\sqrt{y})^2}\bigg],\notag\\
 h_{\phi e}(x)&=\frac{1}{36}\bigg[\frac{x(2x-7)+11}{(x-1)^3}-\frac{6\log x}{(x-1)^4}\bigg],\notag\\
 \tilde h_{\phi e}(x)&=-\frac{1}{36}\bigg[\frac{x(16x-29)+7}{(x-1)^3}-\frac{6x^2(2x-3)\log x}{(x-1)^4}\bigg],\notag\\
 k_{\phi e}(x)&=\frac{1}{x}\tilde h_{\phi e}\Big(\frac{1}{x}\Big)=\frac{1}{36}\bigg[\frac{x(7x-29)+16}{(x-1)^3}+\frac{6(3x-2)\log x}{(x-1)^4}\bigg],\notag\\
 \tilde k_{\phi e}(x)&=\frac{1}{x}h_{\phi e}\Big(\frac{1}{x}\Big)=-\frac{1}{36}\bigg[\frac{x(11x-7)+2}{(x-1)^3}-\frac{6x^3\log x}{(x-1)^4}\bigg]. 
\end{align}

\bibliographystyle{apsrev4-1_mod}
\bibliography{amu}

\end{document}